\documentclass[preprints,article,accept,moreauthors,pdftex]{Definitions/mdpi} 

\extrafloats{100}
\firstpage{1} 
\pubvolume{1}
\issuenum{1}
\articlenumber{0}
\pubyear{2022}
\copyrightyear{2020}
\datereceived{} 
\dateaccepted{} 
\datepublished{} 
\hreflink{https://doi.org/} 

\Title{The AWAKE Run\,2 programme and beyond}

\TitleCitation{Title}

\Author{
Edda~Gschwendtner $^{1,\dagger}$,
Konstantin~Lotov $^{2,\dagger}$,
Patric~Muggli $^{1, 3,\dagger}$,
Matthew~Wing $^{4,\dagger,*}$,
Riccardo~Agnello $^{5}$,
Claudia~Christina~Ahdida $^{1}$,
Maria Carolina Amoedo Goncalves $^{1}$,
Yanis Andrebe $^{5}$,
Oznur Apsimon $^{6, 7}$,
Robert~Apsimon $^{7, 8}$,
Jordan Matias Arnesano $^{1}$,
Anna-Maria Bachmann $^{3}$,
Diego Barrientos $^{1}$,
Fabian Batsch $^{3}$,
Vittorio Bencini $^{1, 9}$,
Michele Bergamaschi $^{3}$,
Patrick Blanchard $^{5}$,
Philip Nicholas Burrows $^{9}$,
Birger~Buttensch{\"o}n $^{10}$,
Allen~Caldwell $^{3}$,
James~Chappell $^{4}$,
Eric~Chevallay $^{1}$,
Moses~Chung $^{11}$,
David Andrew~Cooke $^{4}$,
Heiko~Damerau $^{1}$,
Can~Davut $^{7, 12}$,
Gabor~Demeter $^{13}$,
Amos Christopher~Dexter $^{7, 8}$,
Steffen~Doebert $^{1}$,
Francesa Ann Elverson $^{1}$,
John~Farmer $^{1, 3}$,
Ambrogio~Fasoli $^{5}$,
Valentin~Fedosseev $^{1}$,
Ricardo~Fonseca $^{14, 15}$,
Ivo~Furno $^{5}$,
Spencer~Gessner $^{1, 16}$,
Aleksandr~Gorn $^{2}$,
Eduardo~Granados $^{1}$,
Marcel~Granetzny $^{17}$,
Tim~Graubner $^{18}$,
Olaf~Grulke $^{10, 19}$,
Eloise~Daria~Guran $^{1}$,
Vasyl~Hafych $^{3}$,
Anthony Hartin $^{4}$,
James~Henderson $^{7, 20}$,
Mathias~H{\"u}ther $^{3}$,
Miklos~Kedves $^{13}$,
Fearghus Keeble $^{4}$,
Vadim~Khudiakov $^{21, 22}$,
Seong-Yeol~Kim $^{11, 1}$,
Florian~Kraus $^{18}$,
Michel Krupa $^{1}$,
Thibaut~Lefevre~$^{1}$,
Linbo~Liang $^{7, 12}$,
Shengli Liu $^{23}$,
Nelson~Lopes $^{15}$,
Miguel Martinez Calderon $^{1}$,
Stefano~Mazzoni~$^{1}$,
David~Medina~Godoy $^{1}$,
Joshua~Moody $^{3}$,
Kookjin~Moon $^{11}$,
Pablo Israel~Morales~Guzm\'{a}n $^{3}$,
Mariana~Moreira $^{15}$,
Tatiana~Nechaeva $^{3}$,
Elzbieta~Nowak $^{1}$,
Collette~Pakuza $^{9}$,
Harsha~Panuganti $^{1}$,
Ans~Pardons $^{1}$,
Kevin Pepitone $^{24}$,
Aravinda~Perera $^{7, 6}$,
Jan~Pucek $^{3}$,
Alexander~Pukhov $^{21}$,
Rebecca Louise Ramjiawan $^{1, 9}$,
Stephane~Rey $^{1}$,
Adam~Scaachi $^{4}$,
Oliver~Schmitz $^{17}$,
Eugenio Senes $^{1}$,
Fernando~Silva $^{25}$,
Luis~Silva $^{15}$,
Christine~Stollberg $^{5}$,
Alban~Sublet $^{1}$,
Catherine Swain $^{7, 6}$,
Athanasios Topaloudis $^{1}$,
Nuno~Torrado $^{15}$,
Petr~Tuev $^{2}$,
Marlene Turner $^{1, 26}$,
Francesco Velotti $^{1}$,
Livio Verra $^{3, 1, 27}$,
Victor~Verzilov $^{23}$,
Jorge~Vieira $^{15}$,
Helmut Vincke $^{1}$,
Martin~Weidl $^{28}$,
Carsten~Welsch $^{7, 6}$,
Manfred Wendt $^{1}$,
Peerawan Wiwattananon $^{1}$,
Joseph~Wolfenden $^{7, 6}$,
Benjamin~Woolley $^{1}$,
Samuel Wyler $^{1}$,
Guoxing~Xia $^{7, 12}$,
Vlada Yarygova $^{2}$,
Michael~Zepp $^{17}$,
Giovanni~Zevi~Della~Porta $^{1}$
(The~AWAKE~Collaboration)
}

\AuthorNames{Firstname Lastname, Firstname Lastname and Firstname Lastname}

\address{
$^{1}$ \quad {CERN, Geneva, Switzerland}\\
$^{2}$ \quad {Novosibirsk State University, Novosibirsk, Russia}\\
$^{3}$ \quad {Max Planck Institute for Physics, Munich, Germany}\\
$^{4}$ \quad {UCL, London, UK}\\
$^{5}$ \quad {Ecole Polytechnique Federale de Lausanne (EPFL), Swiss Plasma Center (SPC), Lausanne, Switzerland}\\
$^{6}$ \quad {University of Liverpool, Liverpool, UK}\\
$^{7}$ \quad {Cockcroft Institute, Daresbury, UK}\\
$^{8}$ \quad {Lancaster University, Lancaster, UK}\\
$^{9}$ \quad {John Adams Institute, Oxford University, Oxford, UK}\\
$^{10}$ \quad {Max Planck Institute for Plasma Physics, Greifswald, Germany}\\
$^{11}$ \quad {UNIST, Ulsan, Republic of Korea}\\
$^{12}$ \quad {University of Manchester, Manchester, UK}\\
$^{13}$ \quad {Wigner Research Center for Physics, Budapest, Hungary}\\
$^{14}$ \quad {ISCTE - Instituto Universit\'{e}ario de Lisboa, Portugal}\\
$^{15}$ \quad {GoLP/Instituto de Plasmas e Fus\~{a}o Nuclear, Instituto Superior T\'{e}cnico, Universidade de Lisboa, Lisbon, Portugal}\\
$^{16}$ \quad {SLAC, Menlo Park, CA, USA}\\
$^{17}$ \quad {University of Wisconsin, Madison, Wisconsin, USA}\\
$^{18}$ \quad {Philipps-Universit{\"a}t Marburg, Marburg, Germany}\\
$^{19}$ \quad {Technical University of Denmark, Lyngby, Denmark}\\
$^{20}$ \quad {Accelerator Science and Technology Centre, ASTeC, STFC Daresbury Laboratory, Warrington, UK}\\
$^{21}$ \quad {Heinrich-Heine-Universit{\"a}t D{\"u}sseldorf, D{\"u}sseldorf, Germany}\\
$^{22}$ \quad {Budker Institute of Nuclear Physics SB RAS, Novosibirsk, Russia}\\
$^{23}$ \quad {TRIUMF, Vancouver, Canada}\\
$^{24}$ \quad {Angstrom Laboratory, Department of Physics and Astronomy, Uppsala, Sweden}\\
$^{25}$ \quad {INESC-ID, Instituto Superior Tecnico, Universidade de Lisboa, Lisbon, Portugal}\\
$^{26}$ \quad {LBNL, Berkeley, CA, USA}\\
$^{27}$ \quad {Technical University Munich, Munich, Germany}\\
$^{28}$ \quad {Max Planck Institute for Plasma Physics, Munich, Germany}
}

\corres{Correspondence: m.wing@ucl.ac.uk}

\firstnote{These authors contributed equally to this work.}

\abstract{Plasma wakefield acceleration is a promising technology to reduce the size of particle accelerators.  Use of high energy protons to drive wakefields in plasma has been demonstrated during Run\,1 of the AWAKE programme at CERN.  Protons of energy 400\,GeV drove wakefields that accelerated electrons to 2\,GeV in under 10\,m of plasma.  The AWAKE collaboration is now embarking on Run\,2 with the main aims to demonstrate stable accelerating gradients of 0.5–1\,GV/m, preserve emittance of the electron bunches during acceleration and develop plasma sources scalable to 100s of metres and beyond.  By the end of Run\,2, the AWAKE scheme should be able to provide electron beams for particle physics experiments and several possible experiments have already been evaluated.  This article summarises the programme of AWAKE Run\,2 and how it will be achieved as well as the possible application of the AWAKE scheme to novel particle physics experiments.}


\begin{document}
\section{Introduction}

When a compact particle bunch or laser pulse enters a plasma column, this drive beam disturbs the free plasma electrons which can then set up an oscillatory motion that leads to strong electric fields (``wakefields”) in the direction of the bunch propagation and also transverse to it.  By injecting a witness bunch of charged particles into the correct phase of the plasma electron oscillation, the system acts as a  particle accelerator.
This offers a compelling alternative to conventional microwave radio-frequency acceleration which is limited to accelerating gradients of about 100\,MV/m,  at which point the metallic structures where the particles are accelerated start to break down.  As plasma is already ionised, it does not suffer from this limitation and accelerating gradients many orders of magnitude higher are possible.  As such, plasma wakefield acceleration is a possible solution to developing accelerators of significantly reduced size for high energy particle physics, or indeed for other applications.

The concept of accelerating particles in plasma was first proposed in the 1970s~\cite{prl:43:267}.  The field has undergone significant development since~\cite{prl:54:693,pp:14:055501,rmp:81:1229,rast:9:63,NJP23-031101}, with progress experimentally,  theoretically and in simulation. This has been aided by technology development in high-power lasers and  high-performance computing.  Many experiments using a laser pulse as a driver have shown that wakefields of 10s of GV/m and beyond are sustainable~\cite{nature:377:606,nature:431:535,nature:431:538,nature:431:541}, with electrons accelerated up to 7.8\,GeV in one acceleration stage of 20\,cm of plasma the highest final energy achieved so far~\cite{prl:122:084801}.  Similar accelerating gradients have been achieved when an electron bunch is used as a driver~\cite{nature:445:741,nature:515:92}, with energy gains of 42\,GeV achieved for particles in a single bunch where the head of the bunch drives the wakefields~\cite{nature:445:741} and 9\,GeV per particle achieved for a witness bunch of electrons~\cite{ppcf:58:034017}.  However, the laser pulses and electron bunches both suffer from a low stored energy meaning that multiple acceleration stages~\cite{rast:9:63,prab:13:101301} are being investigated in order to achieve the high energies needed for particle physics experiments.

The possibility to use proton bunches allows the acceleration to take place in one stage given the high stored energy available in some proton accelerators.  The original proposal~\cite{np:5:363} considered, in simulation, TeV protons in bunches of length 100\,$\mu$m which are not currently available.  High energy proton bunches at CERN are typically 10\,cm long, however such bunches can undergo a process called self-modulation (SM) in plasma~\cite{bib:kumar,prl:107:145002,bib:pukhov} in which the long proton bunch is split into a series of microbunches.  These microbunches are regularly spaced and hence can constructively interfere to drive strong wakefields.  The SM process allows the use of proton beams that currently exist in order to develop proton-driven plasma wakefield acceleration into a technology for future particle physics experiments.

The advanced wakefield (AWAKE) experiment at CERN~\cite{ppcf:56:084013,NIMA-829-3,nim:a829:76,bib:muggliready} was developed in order to initially verify the concept of proton-driven plasma wakefield acceleration.  Proton bunches from the super proton synchrotron (SPS), in which each proton has an energy of 400\,GeV, have a total energy of 19\,kJ per bunch.  The bunches are typically 6--12\,cm long and undergo the SM process in a rubidium plasma.  Witness bunches of electrons can then be accelerated in the wakefields driven by the proton microbunches.  Initial experiments were performed in 2016--18 in order to demonstrate the proof of concept (Run\,1).  The scheme is now being developed with a series of experiments (Run\,2) to be performed in this decade.  These will demonstrate it as a usable technology for high energy particle acceleration which already has several potential applications in particle physics.

The outline of the article is as follows.  After this introduction, the highlights of the AWAKE Run\,1 programme are summarised in Section~\ref{sec:run1}.  The physics programme of AWAKE Run\,2 is discussed in Section~\ref{sec:run2} followed by a discussion of the setup for AWAKE Run\,2 in Section~\ref{sec:setup}.  Section~\ref{sec:applications} then summarises the possible particle physics experiments that could be realised after Run\,2 with electrons provided by the AWAKE scheme.  A brief summary is then given in Section~\ref{sec:summary}.

\section{Summary of experimental results from AWAKE Run\,1}
\label{sec:run1}

\begin{figure}[bthp!]
\centering
\includegraphics[width=0.75\textwidth]{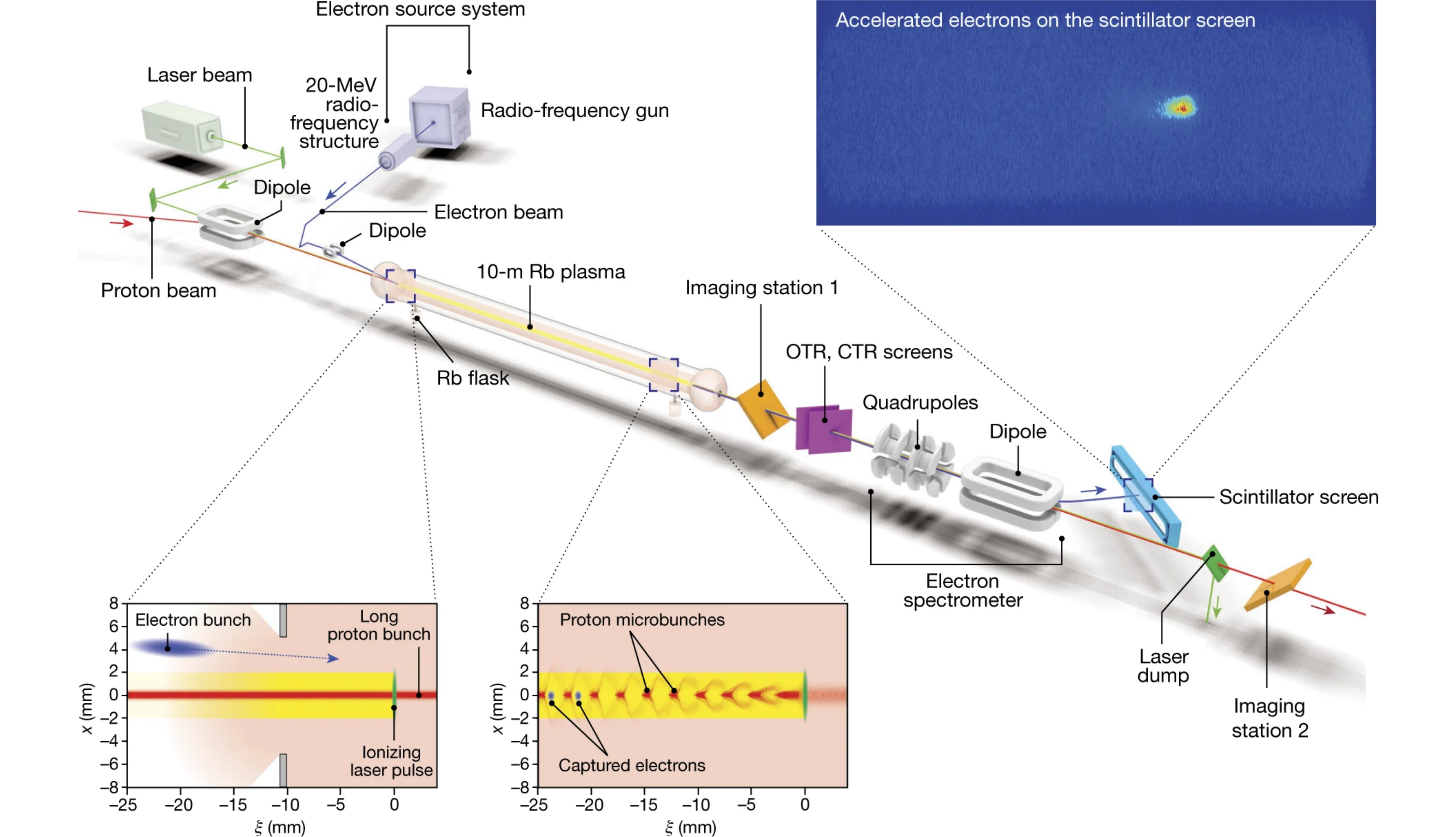}
\caption{Schematic of the AWAKE Run\,1 (2016--18) layout.  The laser and proton beams are merged before entering the plasma source.  A beam of 10--20\,MeV electrons is also merged with the beam line and injected into the entrance of the plasma source.  The plasma source contains rubidium vapour at about 200\,$^\circ$C with precise temperature control over the full 10\,m.  The beams exit the plasma source and a series of diagnostics are used to characterise them.  There are two imaging stations to measure the transverse profile of the proton bunch and screens emitting optical and coherent transition radiation (OTR and CTR) to measure the longitudinal profile of the proton bunch.  Electrons are separated from the protons using a dipole magnet which also induces an energy-dependent spread which is measured on a scintillator screen, imaged by a camera.  Diagrams of the proton bunch self-modulation and electron capture are shown in the bottom left.  A typical image of the accelerated electron bunch as observed on the scintillator screen is shown in the top right.  From Ref.~\cite{bib:nature}.
    }
\label{fig:run1-layout}
\end{figure}

In the first round of experiments~\cite{bib:muggliready}, we have demonstrated the existence of the SM process and the possibility to accelerate electrons in SM-driven wakefields.  We have also observed a number of expected and unexpected characteristics of SM. %
An overview of the experimental setup for Run\,1 is shown in Fig.~\ref{fig:run1-layout}.

\begin{figure}[bthp!]
\centering
\includegraphics[width=0.7\textwidth]{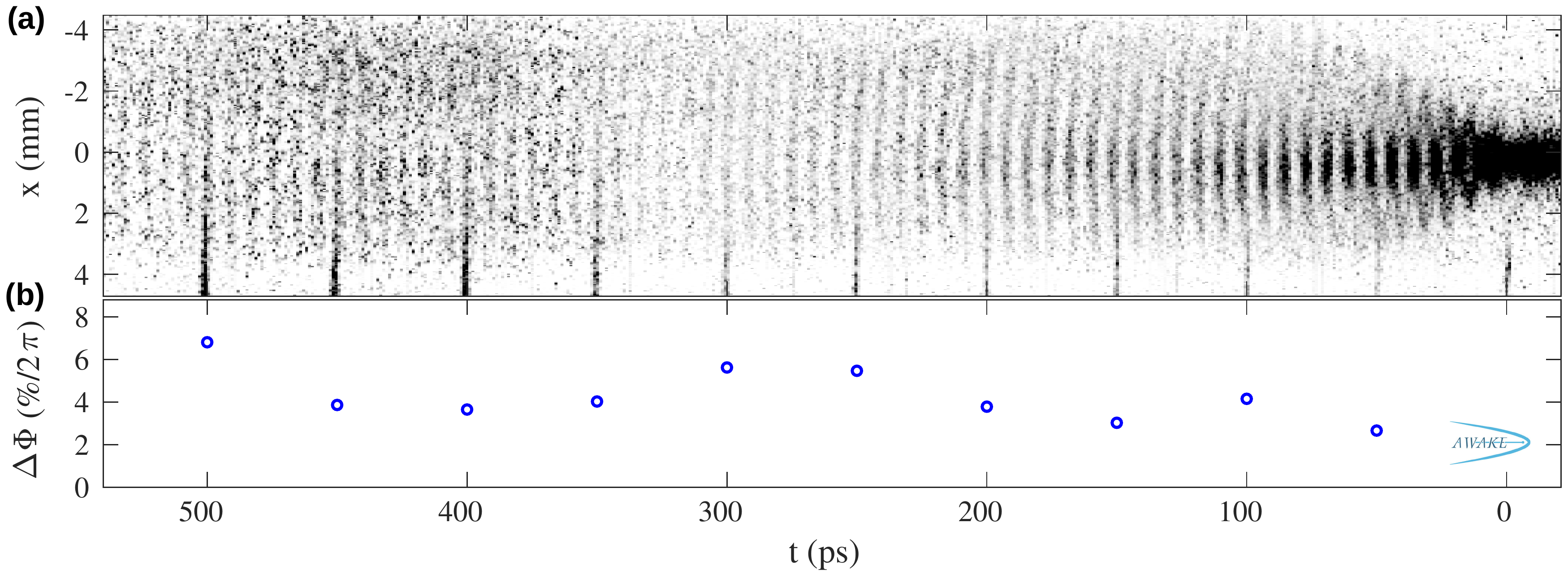}
\caption{
    (a) Time-resolved image of the SM proton bunch with the RIF placed 125\,ps (0.5$\sigma_t$, where $\sigma_t$ is the RMS duration of the proton bunch) ahead of bunch centre (front of the bunch at $t<0$\,ps), and plasma electron density $n_{e0}=1.81\times$10$^{14}$\,cm$^{-3}$ (other parameters in~\cite{bib:fabianprl}). %
    The RIF is at $t=0$\,ps on the image. %
    (b) Relative RMS phase variation $\Delta\Phi$ of the modulated bunches (in \% of 2$\pi$ or of a modulation period) for each set of images acquired every 50\,ps along the bunch and aligned in time using a reference laser pulse signal visible as a vertical line at the bottom of image (a) ($x>2$\,mm). %
    From Ref.~\cite{bib:fabianprl}.}
\label{fig:alongbunch}
\end{figure}

The proton bunch propagates in a plasma created by a relativistic ionisation front (RIF). %
The RIF is the result of the propagation of a short and intense laser pulse in a rubidium vapour~\cite{bib:oz,bib:fabiandensity,bib:gabor, pr:a104:033506}. %
When the RIF is placed within the proton bunch, the part of the bunch behind the RIF travelling in plasma is transformed into a train of microbunches. This is shown in Fig.~\ref{fig:alongbunch}~(a) where a clear periodic charge density structure at $t>0$\,ps is observed. %
The front ($t<0$\,ps on \mbox{Fig.~\ref{fig:alongbunch}(a)}) is unaffected. %
The period of the train or the modulation frequency is determined by the plasma electron frequency,  $f_{pe}$~\cite{bib:karl}, measured over one order of magnitude in plasma electron density $n_{e0}$: $f_{pe}\propto \sqrt{n_{e0}}$~\footnote{The electron plasma frequency in a plasma with electron density $n_{e0}$ is: $\omega_{pe}=\left(\frac{n_{e0}e^2}{\varepsilon_0m_e}\right)^{1/2}$, $f_{pe}=\omega_{pe}/2\pi$, where $e$ is the elementary electric charge, $\varepsilon_0$ is the permittivity of free space and $m_e$ is the electron mass.}. %
The train formation is a transverse process; protons between microbunches leave the bunch axis and form an expanding halo. %
The halo radius measured 10\,m downstream from the plasma exit indicates that protons gained radial momentum from transverse wakefields with amplitudes reaching hundreds of MV/m~\cite{bib:marleneprl}. %
This amplitude exceeds their initial amplitude driven at the RIF at the plasma entrance ($<$10\,MV/m). %
This indicates that wakefields grow along the plasma, whereas the increase in radius reached by halo protons along the bunch shows that they also grow along the bunch, with both growths expected \cite{NIMA-829-3}. %
Correspondingly, large longitudinal wakefields lead to a 2\,GeV energy gain of externally injected 19 MeV test electrons~\cite{bib:nature,bib:marleneroyal}. %
Acceleration experiments also suggest that wakefields may break in the back of the bunch, due to the large amplitude of the wakefields and to the finite radial extent of the plasma~\cite{bib:james,PPCF63-055002-halo}.
Combined halo radius and acceleration results in experiment and simulations show that the SM process saturates a distance between three and five metres along the plasma~\cite{bib:marleneprab}. %

Numerical simulation results suggest that injection and acceleration of electrons are ineffective during the SM growth.
If the electrons are injected along the proton beam axis, they are defocused in the region of longitudinally varying plasma density near the entrance to the plasma section~\cite{PoP21-123116, NIMA-829-3, PoP25-063108}. 
If the electrons are injected at a small angle, they cross the transverse boundary of the plasma column which reflects or scatters most of the injected charge~\cite{PPCF61-104004}. 
This is the likely reason why the measured accelerated charge \cite{bib:nature,bib:marleneprab} is significantly smaller than that simulated without the effect of boundary crossing~\cite{NIMA-829-3}.

The phase velocity of the wakefields during SM is less than the speed of light $c$~\cite{bib:pukhov,prl:107:145002}, which prevents electrons from being accelerated to high energies \cite{IPAC14-1537}. 
This also causes microbunch destruction at late stages of SM development~\cite{PoP18-024501, PoP22-103110}. 
Fortunately, the phase velocity of the wakefields can be influenced by plasma density gradients along the beam trajectory \cite{NIMA-829-3}. %
It also changes the number of protons remaining in the microbunches and the length of the train: more charge and a longer train with positive density gradients, as demonstrated experimentally~\cite{bib:falk} and detailed in simulations~\cite{bib:pablo}. %

Experimental and simulation results also reveal that with a density gradient the modulation frequency is not unique and varies radially across the time-resolved charge distribution of the train and halo observed 3.5\,m from the exit of the plasma~\cite{bib:pablo}. %
The structure of the distribution confirms that protons forming the halo left the wakefields over the first few metres of plasma, their distribution carrying the plasma frequency near the plasma entrance. %
The microbunch train, having travelled through the entire plasma, carries the frequency at the end of the plasma. %
We observe this with negative, linear density gradients~\cite{bib:falk} with which we also observe the largest frequency variations, with shorter trains, better able to adjust their modulation frequency to the local plasma frequency. %

The reproducibility of the accelerating structure in the plasma is essential for the controlled acceleration of an externally-injected electron bunch. %
The electron bunch must be placed at the proper phase within the wakefields. %
That is, at a position within the train where wakefields have reached their maximum, and more precisely, within a fraction of a period of the wakefields, where fields are accelerating and focusing. %
The precise location is determined, for example, by loading of the wakefields for minimisation of energy spread and for emittance preservation~\cite{bib:veronica}. %
While we do not measure the reproducibility of the wakefields from event to event, we demonstrated that when the RIF provides initial wakefields with sufficient amplitude, the phase of the bunch modulation with respect to the RIF is reproducible, despite variation of the incoming bunch parameters~\cite{bib:fabianprl}. %
That is, the SM process is seeded, i.e., reproducible and driven away from its instability (SMI) regime~\cite{bib:kumar}. %
In this seeded mode, we measure RMS variations of the modulation phase smaller than 8\% of a modulation period (Fig.~\ref{fig:alongbunch}~(b)) all along the bunch train. %
We observe the instability when the RIF is placed further than $\approx2\sigma_t$ ahead of the centre of the bunch with RMS duration $\sigma_t$ (typically $\cong250$\,ps). %
When we place the RIF $\gg2\sigma_t$ ahead of the centre of the bunch, the bunch propagates in a pre-formed plasma whose density is decaying in time because of radial expansion and recombination. %
Recording the modulation frequency as a function of RIF timing provides a measurement of plasma density as a function of time after ionisation~\cite{bib:gessner}. %

While the SM process is the lowest-order, symmetric mode of interaction between the long incoming bunch and the plasma, signs of the non-axi-symmetric mode, the hose instability~\cite{bib:whittum,thesis:huether}, were also observed. %
However, this mode was only observed at low plasma densities, $n_{e0}\le0.5\times10^{14}\,{\rm cm}^{-3}$, much lower than those that led to significant energy gain, $n_{e0}>1.8\times10^{14}\,{\rm cm}^{-3}$. %

The above results, in particular the occurrence and the saturation of the SM process of the 400\,GeV proton bunch over a distance less than 10\,m of plasma, with its phase reproducibility, the possibility to accelerate electrons in the wakefields, the absence of hosing instability and the generally excellent agreement between experimental and simulation results~\cite{bib:james,bib:gorn,bib:marleneprab,bib:pablo}, allow for the planning of the next experiments~\cite{bib:mugglirun2}. %
These will be conducted in a number of steps geared towards experiments with two plasmas aimed at producing a multi-GeV electron bunch with charge, emittance and relative energy spread sufficient for the applications described in this manuscript. 

\section{The AWAKE Run\,2 physics programme}
\label{sec:run2}


The Run\,2 physics programme is driven mostly by the long-term goals presented in this paper: producing a high-energy electron bunch with quality sufficient for high-energy or particle physics applications. %
Run\,2 will again use proton bunches from the SPS. %
The main difference with Run\,1 is the use of two plasma sources, one for SM and one for acceleration, thereby allowing for on-axis injection of the electron bunch into the accelerator plasma and for better control of parameters~\cite{run2a-plan}. %

The first part of Run\,2 focuses on the self-modulator, i.e., on the generation of the self-modulated proton bunch to drive the accelerator. %
The second part of Run\,2 focuses on the accelerator, i.e., on external injection of the electron bunch and on scaling of its energy gain to higher energies. %

We determined in Run\,1 that, as predicted by numerical simulations~\cite{NIMA-829-3,PoP21-083107}, the SM process saturates over a distance of 3--5\,m~\cite{bib:marleneprab}. %
Experiments with two plasmas will thus include a 10\,m-long self-modulator plasma, followed by a 10\,m-long accelerator plasma. %
In the current plan, the two plasma sources will be based on laser ionisation of a Rb vapour~\cite{bib:oz}. %
These are the only sources known so far that  provide a plasma density step in the self-modulator~\cite{bib:plyushchev} and the desired density uniformity~\cite{PoP20-013102} in the accelerator. %

    \subsection{Self-modulator}
    The self-modulator will have two new features: the ability of seeding the SM process using an electron bunch and the ability of imposing a plasma density step. %
    These two features will be first tested independently, and then together. %

        \subsubsection{Electron-bunch seeding}
        A major result of Run\,1 was the demonstration of the seeding of the SM process using a RIF~\cite{bib:karl}. %
        However, this seeding method leaves the front of the bunch, ahead of the RIF, un-modulated. %
        Since AWAKE long-term plans call for an accelerator plasma tens to hundreds of metres long, this plasma will have to be pre-formed. %
        The laser ionisation process of Run\,1 does not scale to such long plasma because of energy depletion of the laser pulse and because of the focusing geometry. %
        In this preformed plasma, the un-modulated front of the bunch could experience SMI in the accelerator plasma. %
        The wakefields driven by this front SMI could interfere with the self-modulated back of the bunch and with the acceleration process. %
        
        The SM process can also in principle be seeded by a preceding driver of wakefields, such as an electron bunch or a laser pulse. %
        In this case, the entire proton bunch becomes self-modulated and the possible issue with the un-modulated front would be avoided. %
        
        The programme thus consists of demonstrating that the SM process can indeed be seeded by the electron bunch available in the Run\,1 experimental setup. %
        The method to be used is similar as that of Run\,1~\cite{bib:fabianprl}, i.e., determining the timing of microbunches appearing along the bunch with respect to the time of the electron bunch, after 10\,m of plasma. %
        
        \subsubsection{Plasma density step}
        \label{sec:density-step}
         Numerical simulation results suggest that in a plasma with constant density along the beam path, the continuous evolution of the bunch train and wakefields leads to a decay of the amplitude of wakefields after their saturation~\cite{PoP18-024501,PoP22-103110}. %
         These results also suggest that when applying a density step, some distance into the plasma, within the growth of the SM process, wakefields maintain a near-saturation amplitude for a long distance along the plasma. %
         Figure~\ref{fig:field_wwo_step_gap} illustrates how the density step changes the wakefield amplitude in the SM and acceleration plasma sections. %
         These simulations are performed in the axi-symmetric geometry with the quasi-static code LCODE \cite{NIMA-829-350, PRST-AB6-061301}. %
         The parameters of the density step were optimised for the strongest wakefield at $z=20$\,m with no gap between the sections~\cite{PPCF62-115025}. %
         The density step is seen to strongly increase the wakefield at the acceleration stage even in the presence of a 1\,m gap. %

         The AWAKE plasma source is based on a rubidium vapour, along which a uniform temperature is imposed to obtain a correspondingly uniform vapour density. %
         Laser-pulse ionisation then turns this uniform vapour density into an equally uniform plasma density~\cite{bib:karl}. %
         One can therefore simply impose a temperature step along the column to obtain the corresponding plasma density step~\cite{bib:plyushchev}. %
         Measurements of the effect of the plasma density step on the amplitude of the wakefields will include effects of the size and shape of the bunch halo formed by defocused protons, measurements of plasma light signals, and measurements of electron acceleration. %


    \subsection{Accelerator}
\begin{figure}[H]
        \centering
        \includegraphics[width=0.6\textwidth]{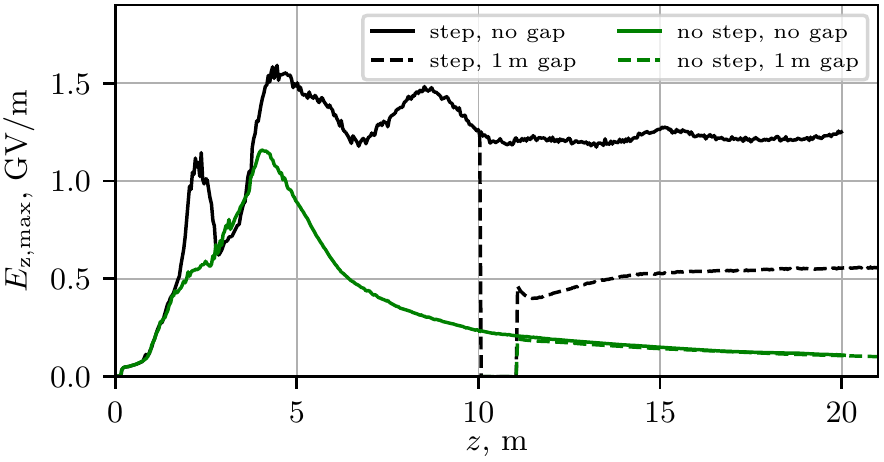}
        \caption{The amplitude of the excited wakefield $E_\text{z,\,max} (z)$ in the uniform plasma and in plasma with the optimised density step with and without a 1\,m gap between SM and acceleration plasma sections. The SM process is seeded by an electron bunch. The density step is the linear growth of the plasma density from $7 \times 10^{14}\,\text{cm}^{-3}$ to $7.21 \times 10^{14}\,\text{cm}^{-3}$ at the interval between $z=0.8$\,m and $z=2.8$\,m.}
        \label{fig:field_wwo_step_gap}
        \end{figure}

    The length of the accelerator plasma is 10\,m for the first experiments. %
    This is much longer than the distance it takes for the amplitude of wakefields to settle to steady values after the injection point (see Fig.~\ref{fig:field_wwo_step_gap}). 
    This distance is $\sim$2\,m and results from the transverse evolution of the proton bunch in the vacuum gap between the two plasmas. %
    The plasma is thus long enough for the expected energy gain to be in the multi-GeV range along the ensuing length of plasma where the driving of the wakefields is stationary. %

        \subsubsection{External injection}
        In Run\,1, acceleration was obtained with an off-axis injection geometry~\cite{NIMA-829-3}. %
        This geometry was chosen to avoid defocusing of the injected electrons in the density ramp located at the entrance of the plasma~\cite{PoP21-123116,PoP25-063108}.
        In this region, the yet un-modulated proton bunch drives transverse fields which are focusing for its own positive-charge sign, but defocusing for injected electrons. %
        A scheme that avoids these issues is to inject electrons on axis after the SM process has saturated. %
        The parameters of the electron bunch must be such that it reaches high energies, low final relative energy spread and preserves its incoming emittance. %
        The SM process does not lead to blow-out of plasma electrons from the accelerating structure because the resonant wave drive stops when the plasma wave becomes nonlinear and its period elongates~\cite{PoP20-083119}.
        The initial electron bunch density $n_{b0}$ must therefore exceed the plasma electron density: $n_{b0}\gg n_{e0}$ to reach blow-out. %
        Blow-out of plasma electrons is necessary for the focusing force of the plasma to become that of the pure ion column, with strength increasing linearly with radius~\cite{PRA44-6189}. %
        The normalised emittance $\epsilon_N$ and focus RMS size of the electron bunch at the plasma entrance $\sigma_{r0}$ must be adjusted to satisfy matching to the ion column focusing force:
        
$$\frac{\gamma_0n_{e0}}{\epsilon_N^2}\sigma_{r0}^4=\frac{2\varepsilon_0m_ec^2}{e^2}.$$ %
        For typical parameters ($\epsilon_N=2$\,mm\,mrad, $n_{e0}=7\times10^{14}\,\text{cm}^{-3}$), electrons must have a large enough energy or relativistic factor $\gamma_0$ to be focused to the small transverse size ($\sigma_{r0}=5.65\,\mu$m, $\gamma_0\cong300$). %
        In addition, the bunch length and timing in the wakefields must be optimised to load wakefields to minimise the final relative energy spread $\Delta E/E$. %
            
        An example set of parameters was developed using a toy-model for the proton bunch~\cite{bib:veronica}. %
        This example shows that with parameters satisfying the above conditions, about 70\% of the initial 100\,pC bunch charge preserve their emittance and reach 1.67\,GeV/c over 4\,m of plasma with $\Delta E/E\cong1\%$ (core). %

        \subsubsection{Scalable plasma sources}
        Assuming the success of experiments on electron injection into the wakefields of a 10\,m-long plasma, energy gain suitable for applications can in principle be achieved by extending the length of the accelerator plasma. %
        However, the distance over which laser ionisation can occur is limited by depletion of the energy of the laser pulse and by the focusing geometry. %
        We are therefore developing other plasma sources that do not suffer from length limitations: direct-current electrical discharge in noble gases~\cite{Torrado2022} and helicon argon plasma~\cite{bib:helicon}. %
            
The direct-current discharge plasma source (DPS) uses a short pulse ($\sim 10\,\mu$s) high-current pulse ($\sim$\,kA) through the length of a glass tube, filled with a high atomic number noble gas at low pressure ($\sim$\,10\,Pa)~\cite{Torrado2022}.  This follows a fast-rising high-voltage pulse ($50 - 100$\,kV) able to ignite long tubes ($L>5$\,m) into uniform plasma densities.  We are currently developing a 10\,m long plasma source consisting of a double discharge from a mid length common cathode to two anodes in the extremities. Operation at the plasma densities relevant for AWAKE has been demonstrated. We are currently engaged in demonstrating the ability of the plasma source to reach the required plasma density uniformity. With this source, the length scalability can potentially be reached by stacking together multiple plasma sections (with lengths of a few metres to a few tens of metres) using common cathodes and anodes.

Helicon plasma belong to the class of magnetised wave heated plasmas~\cite{helicon1}. They consist of a dielectric vacuum vessel and an external wave excitation antenna, which is powered by radio  frequencies. The excited helicon waves heat the plasma and have been demonstrated to generate discharges with high plasma densities~\cite{helicon2}.
            Its length can thus in principle be extended over long lengths by stacking cells. %
            Measurements show that the plasma density typical of AWAKE ($n_{e0}=7\times10^{14}\,{\rm cm}^{-3}$) can be reached~\cite{bib:helicon}. %
            The challenge is to demonstrate that the required plasma density uniformity can also be reached. %
This demonstration requires highly accurate and localised plasma density measurements. Specific diagnostics such as Thomson scattering~\cite{epfl-paper} and optical emission spectroscopy are being developed. 


\section{Overview of AWAKE Run\,2 setup}
\label{sec:setup}

\begin{figure}[htb]
        \centering
        \includegraphics[width=0.7\textwidth]{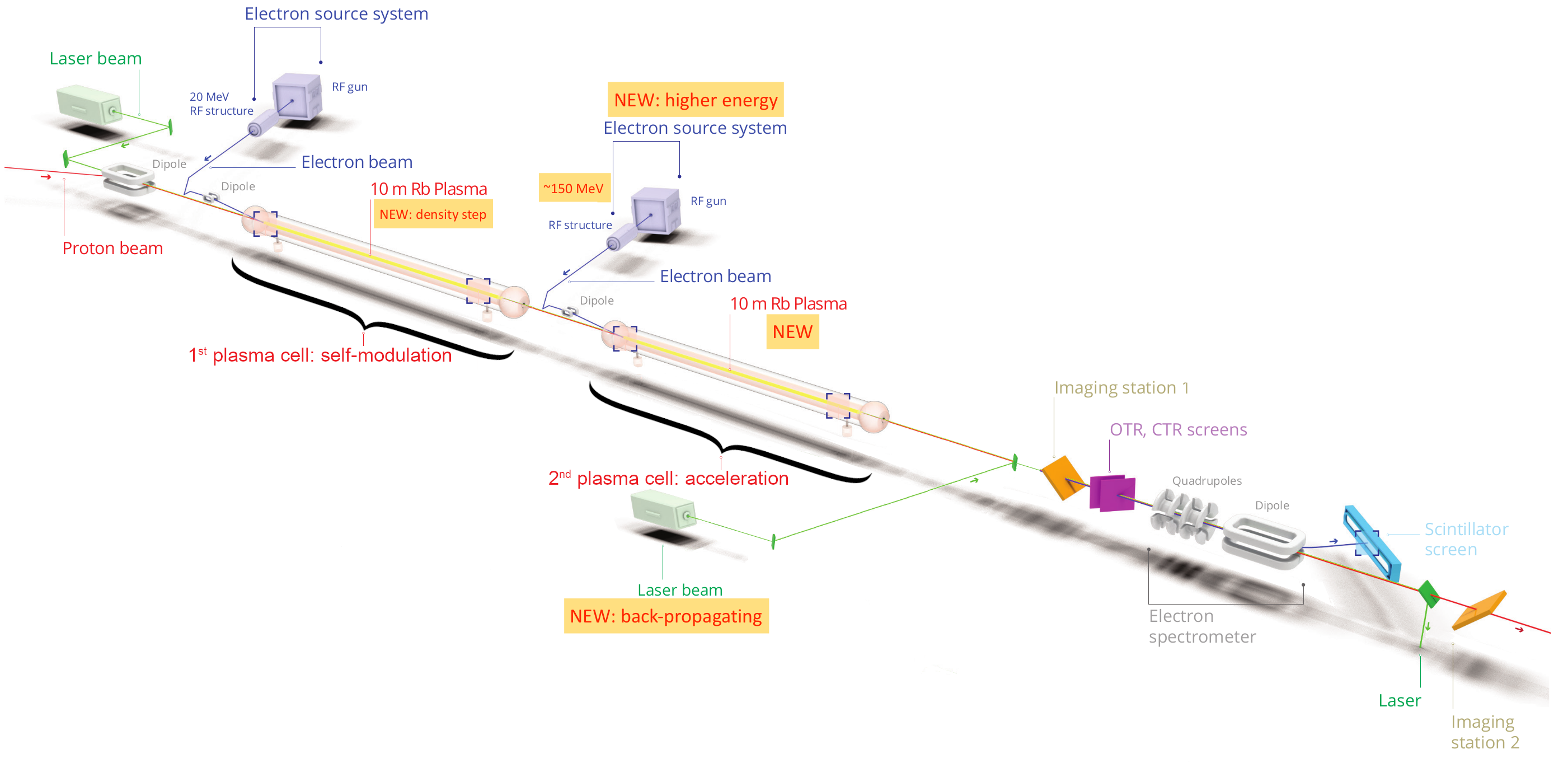}
        \caption{Layout of AWAKE Run\,2.}
        \label{fig:layout_Run2}
        \end{figure}

The AWAKE Run\,2 scheme including the two plasma sources, i.e., a self-modulator and an accelerator, and a new electron beam system is shown in Fig.~\ref{fig:layout_Run2}.

The AWAKE Run\,2 programme is subdivided into four phases Run\,2a, 2b, 2c and 2d, following the physics programme described above. 

The current AWAKE experiment is installed upstream of the previous CERN neutrinos to Gran Sasso (CNGS) facility~\cite{bib:cngs}. The $\sim$\,100 m long CNGS target cavern, which currently houses the CNGS target and secondary beam line while their activation levels decay, is separated from the AWAKE experiment by a shielding wall.

Phases Run\,2a and Run\,2b will be carried out in the existing AWAKE facility, having started in 2021 and are foreseen to continue until 2024. 
For Run\,2c and Run\,2d, however, the CNGS target area needs to be dismantled; 
around three years are required for the removal of the CNGS target area and the installation and commissioning of the additional equipment for AWAKE Run\,2c, where AWAKE takes advantage of the general 1--2 years shutdown period of the CERN injector complex in 2026/27. It is planned to start AWAKE Run\,2c with first protons after LS3 in 2028. 

\subsection{AWAKE Run\,2a}
The Run\,2a experiments, focusing on electron bunch seeding of the SM process, use the same infrastructure as that of AWAKE Run\,1; bunches of protons are extracted from the CERN SPS and each consists of $3\times 10^{11}$\,protons, each with energy 400\,GeV. The bunch length can be adjusted between 6 and 12\,cm and the bunch is focused at the plasma entrance to a transverse RMS size of $\sigma_r \approx 0.2\,$mm. 
The plasma is 10\,m long, has a radius of approximately 1\,mm and a density uniformity better than 0.1\%~\cite{bib:fabiandensity}. 
To create the plasma, rubidium is evaporated in a heat exchanger and the outermost electron of each rubidium atom is ionised with a laser pulse with a pulse length of 120\,fs and a pulse energy of $\leq$\,450\,mJ. The vapour (and thus also the plasma) density is controlled by the temperature of the source and is adjustable between 0.5 and 10 $\times 10^{14}$\,atoms/cm$^3$. 

In AWAKE Run\,1, the relativistic ionisation front of the laser was also  used to seed the self-modulation process by placing it close to the centre of the proton bunch. In AWAKE Run\,2a the laser pulse is placed in front of the  proton bunch so that the entire proton bunch interacts with plasma. 

In AWAKE Run\,1,  electron acceleration in the proton-driven plasma wakefield was demonstrated with externally injected 10--20\,MeV electrons~\cite{bib:nature}. These 100--600\,pC electron bunches have a duration of $\sigma_t \geq4$\,ps and are produced in a RF photo injector based on an S-band structure. 
In AWAKE Run\,2a, these electrons are used to seed the proton bunch SM.

\subsection{AWAKE Run\,2b}
\label{sec:density-step2}
In the Run\,2b experiments, the effect of the plasma density step on the SM process will be measured. This requires a new vapour source and corresponding new diagnostics. The new vapour source is in its design phase and will be exchanged with the current source for Run\,2b. It includes additional observation ports in order to diagnose the electron plasma density that sustains wakefields. The experimental programme will focus on direct measurements of the plasma wakefields. To this end, different diagnostics are currently being evaluated (e.g.\ THz shadowgraphy diagnostics and plasma light diagnostics). 

\subsection{AWAKE Run\,2c}
The CNGS cavern needs to be emptied in order to house the baseline AWAKE Run\,2c and Run\,2d experiments, which  includes the second electron source, beam line, klystron system and the second vapour source. 

In order to be able to integrate the entire Run\,2c experiment in the AWAKE facility, the first plasma cell will be shifted by around 40\,m downstream of its current location and consequently the new equipment will also be accordingly moved downstream. 
This change also needs some downstream shifting of the proton beam line final dipole magnets, however, despite challenging aperture constraints, no extra magnets are required. 
The second electron source needs to deliver electrons with 150\,MeV energy, bunch charge of a few 100\,pC, beam emittance of\,2 mm\,mrad and a short bunch length of 200\,fs duration. The baseline proposal is a novel RF gun and two X-band structures for velocity bunching and acceleration. 
A prototype system is currently being developed %
in order to demonstrate the required beam parameters for AWAKE and to study the mechanical and integration aspects in the AWAKE facility. 

The beam line design for the new 150\,MeV electron beam from the electron source to the plasma source is very challenging, given the tight beam specifications~\cite{Ramjiawan:2021gpy}: the beam must be matched to the plasma at the plasma merging point, with a RMS beam size satisfying $\sigma_{x,y} = \sqrt{0.00487\epsilon_{x,y}}$, zero dispersion and $\alpha_{x,y}= 0$, for emittance, $\epsilon_{x,y}$, and Twiss parameter, $\alpha_{x,y}$, where 0.00487 is the required $\beta$ function in metres. In addition, the gap between the two vapour sources must be as short as possible~\cite{IPAC16-2557}, i.e., $\le 1\,$m. Also considering integration limits from the width of the tunnel, the baseline proposal of the electron transfer line is a dogleg design. Studies on the injection tolerances of the proton and electron beam are currently ongoing and are key to controlled plasma wakefield acceleration. 

The accelerator plasma source will have a length of about 10\,m and will be based on the laser ionised Rb vapour source (as used in AWAKE Run\,1 and in the first vapour source).  The laser beam for the ionisation in the second vapour source will be injected from its downstream end, counter-propagating to the proton beam. Although the same laser as for the first vapour source can be used in Run\,2c by splitting its output beam on two branches, additional laser transport lines and a compressor chamber need to be integrated.

\subsection{AWAKE Run\,2d}

Once Run\,2c has demonstrated electron acceleration to high energies while controlling beam quality, the second plasma source can be exchanged with a different plasma technology.  These sources will be scalable to long distances in order to accelerate electrons to energies of several 10s of GeV and beyond, allowing for the first particle physics applications. 

As discussed in the previous section, the plasma technologies currently under study at CERN are helicon plasma sources and discharge plasma sources. 

With the CNGS target cavern fully dismantled, there is enough space available to install a plasma source of 10s of metres in length. Therefore infrastructure changes for Run\,2d concern mainly the different services such as powering, cooling, etc. needed for the plasma source. 

First studies also show that enough space is available for a prototype fixed-target experimental 
setup, allowing the first particle physics experiments to be conducted with electrons accelerated from AWAKE.

\section{Particle physics applications of AWAKE}
\label{sec:applications}

As outlined in the previous sections, the AWAKE scheme aims to develop an acceleration technology that can be used to provide beams for particle physics experiments.  The ultimate goal for novel acceleration schemes is to provide the technology for a high energy, high luminosity, linear electron--positron collider with centre-of-mass energies $O({\rm TeV})$ and luminosities $10^{34}$\,cm$^{-2}$\,s$^{-1}$. In principle, though, any experiment that requires a source of bunched high energy electrons could utilise the AWAKE scheme.  Initial application focuses on experiments with less challenging beam parameters than a linear $e^+e^-$ collider, although still having a strong and novel particle physics case.

By the conclusion of Run\,2, the AWAKE collaboration should have demonstrated acceleration of electrons with stable GV/m gradients.  Scalable plasma sources should have been developed that can be extendable up to even kilometres in length.  The acceleration process should preserve the beam quality resulting in bunches with transverse emittance of below 10\,mm\,mrad.  With these developments, using proton bunches from the SPS, acceleration of electrons to 10s of GeV, and even up to $\sim$\,200\,GeV, should be possible~\cite{Lotov:2021nob}.  Use of the LHC protons with energy 7\,TeV would enable acceleration of electrons up to about 6\,TeV~\cite{bib:caldwell}.  A limitation of the current proton drivers is their repetition rate and hence the luminosity of any application of the AWAKE scheme.  Given this, high energy applications are considered and also those where the luminosity is less critical.

First ideas for particle physics experiments based on electron bunches from the AWAKE scheme were proposed in previous papers~\cite{Wing:rsta,Caldwell:2018atq}.  The applications are discussed below, along with extensions and new ideas that have been developed since.  The first application would be to use a high energy electron beam impinging on a target in order to search for new phenomena related to dark matter, see Section~\ref{sec:dark-photons}.  Another potential first application is collision of an electron bunch with a high-power laser pulse to investigate strong-field QED, see Section~\ref{sec:sfqed}.  Both of these experiments would require electrons of 10s of GeV, although higher energies could be considered.  An electron--proton collider would be the potential first use of the AWAKE scheme for a high-energy collider.  There is a physics case for such a collider with electrons of 10s of GeV, and up to the TeV scale, when in collision with protons from the LHC.  Each has a strong particle physics case but with less strict demands on beam quality than an $e^+e^-$ collider, see Section~\ref{sec:ep}.  Finally, a new development is consideration of the proton beam at Brookhaven National Laboratory (BNL) as the wakefield driver to develop a compact electron injector which is discussed in Section~\ref{sec:eic}.

\subsection{A beam-dump experiment for dark photon searches}
\label{sec:dark-photons}

Dark photons are postulated particles~\cite{jetp:56:502,pl:b136:279,pl:b166:196} which could provide the link to a dark or hidden sector of particles.  This hidden sector could explain a number of issues in particle physics, not least of which is that they are candidates for dark matter which is expected to make up about 80\% of known matter in the Universe.  Dark photons are expected to have low masses (sub-GeV)~\cite{pl:b513:119,pr:d79:015014} and couple only weakly to Standard Model particles and so would have not been seen in previous experiments.  The dark photon, labelled $A^\prime$, is a light vector boson which results from a spontaneously broken new gauge symmetry and kinetically mixes with the photon and couples to the electromagnetic current with strength $\epsilon \ll 1$.  

A common approach to search for dark photons is through the interaction of an electron with a target in which the dark photon is produced and subsequently decays.  Many experiments, current and proposed~\cite{arXiv:1707.04591}, are searching for dark photons and other feebly interacting particles using  
electrons impinging on a target.  The initial electron beam energy varies, although only the NA64 experiment at CERN~\cite{pr:d89:075008,arxiv:1312.3309,prl:118:011802} has access to high energy electrons ($O(100\,{\rm GeV}$)) with other experiments limited to below ($O(10\,{\rm GeV}$)).  A limitation of experiments is the rate of electrons on target which in the case of NA64 is about $10^6$\,electrons on target per 
second as they are produced in secondary interactions of the SPS proton beam.  Given the limitations of the number of electrons on target for a high energy beam, the AWAKE acceleration scheme could enable an experiment to extend the search for  dark photons as the number of electrons is expected to be several orders of magnitude higher.  Assuming~\cite{awake++} a bunch of $5 \times 10^9$ electrons, each of 50\,GeV in energy, and a running period of 3\,months gives $10^{16}$ electrons on a target of centimetre transverse size.  As the AWAKE scheme produces bunches of electrons, an experiment based on this will run as a beam-dump experiment in which electrons are absorbed in a target and a search for dark photons decaying to an $e^+e^-$ pair is performed.  This is in contrast to other fixed-target experiments in which single electrons impact on a target and other decay channels can be searched for such as dark photons decaying to a pair of dark matter particles which do not leave deposits in the detectors and so have a signal of missing energy.

\begin{figure}[bthp!]
\centering
\includegraphics[width=0.7\textwidth]{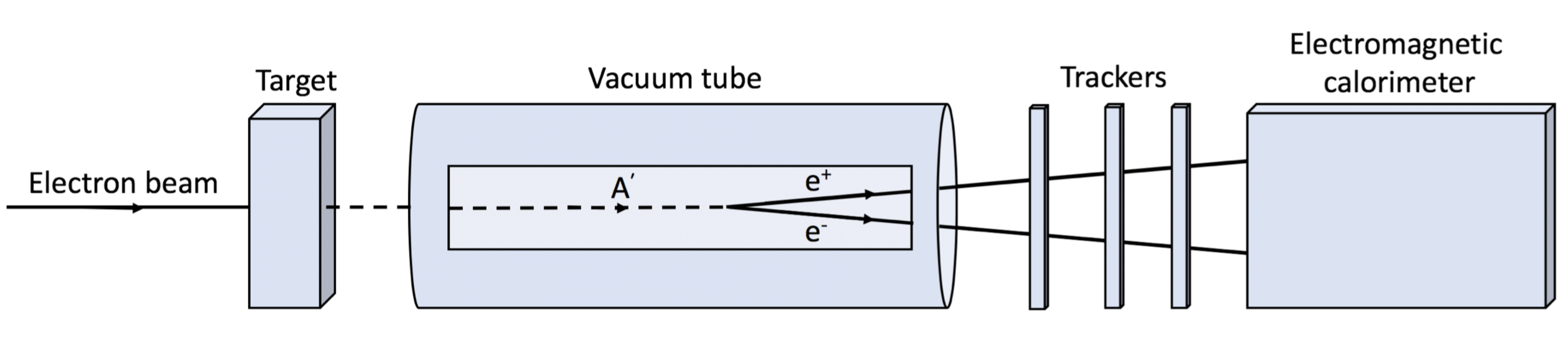}
\caption{Schematic layout of an experiment to search for dark photons.  In the AWAKE scheme, a bunch of  electrons enters from the left and impacts on a target of $O(1\,{\rm m})$ in length.  A produced dark photon travels through a vacuum tube of length $O(10\,{\rm m})$ in which it decays to an $e^+e^-$ pair which are then measured in a detector system such as a tracking detector and calorimeter.
}
\label{fig:dark_photons_layout}
\end{figure}

\begin{figure}[h]
\centering
\includegraphics[width=0.36\textwidth]{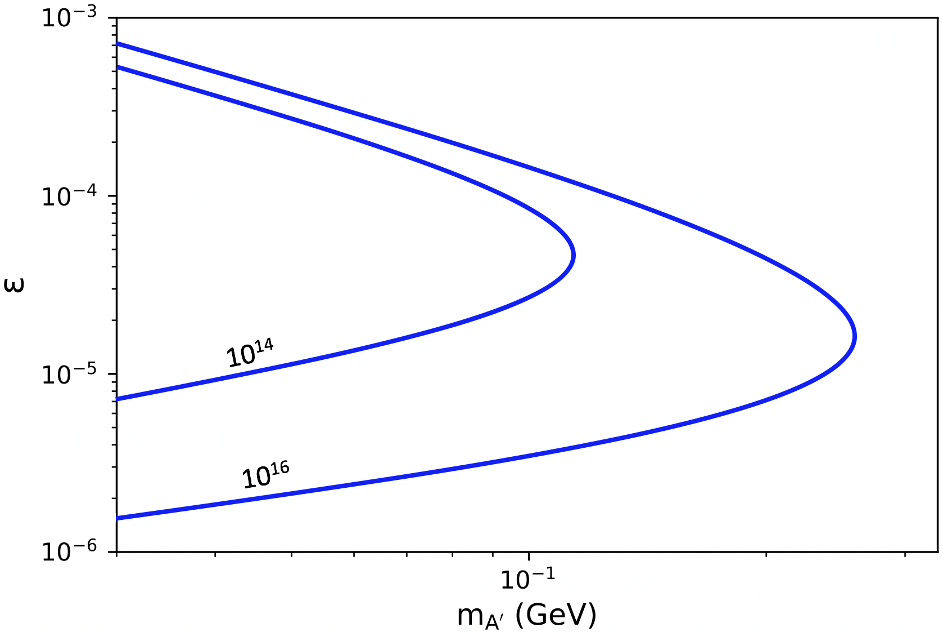}
\includegraphics[width=0.36\textwidth]{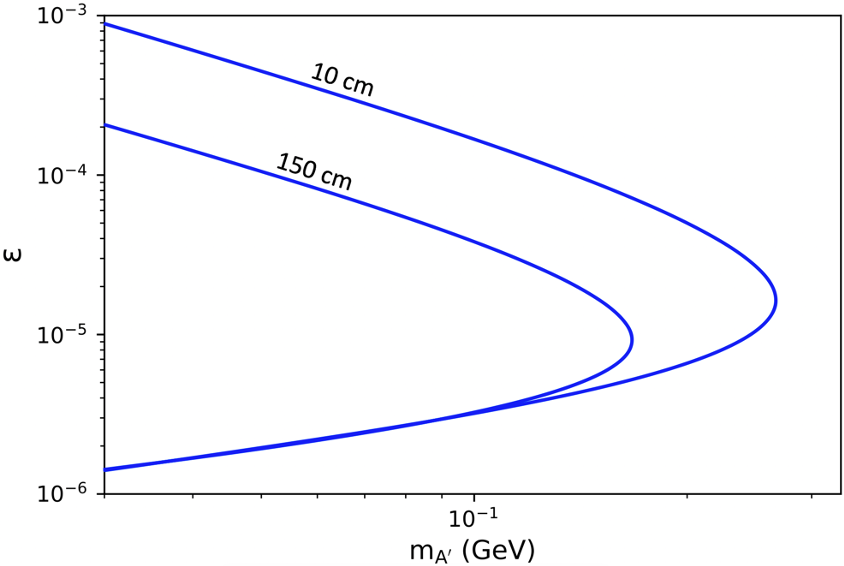}
\caption{Sensitivity to dark photon production shown for the coupling strength, $\epsilon$, and mass, $m_{A^\prime}$.  The varied parameters of the proposed beam-dump experiment are (left) the number of electrons on target and (right) the thickness of the solid metal target which the electrons hit.  The initial electron energy is assumed to be 50\,GeV.}
\label{fig:dark_photons_param}
\end{figure}

The potential production of dark photons is sensitive to the experimental setup.  Using initial electrons of 50\,GeV, an experimental setup, as shown in Fig.~\ref{fig:dark_photons_layout}, has been simulated using GEANT4~\cite{geant4-1,geant4-2,geant4-3}.  To characterise the performance of the experiment, the sensitivity to the coupling strength, $\epsilon$, and mass, $m_{A^\prime}$, is considered and usually represented in a plot of the two.  Examples are shown in Fig.~\ref{fig:dark_photons_param} in which the sensitivity is shown for the number of electrons on target and the thickness of the target.  These results show the value of having as many electrons on target where the sensitivity to both $\epsilon$ and $m_{A^\prime}$ is increased with an increasing number of electrons.  
They also show that the sensitivity is reduced with increasing target thickness, however, having a thicker target is necessary in order 
to keep the background rate under control.     

\begin{figure}[bthp!]
\centering
\includegraphics[trim={5cm 0cm 6cm 1.5cm},clip,width=0.5\textwidth]{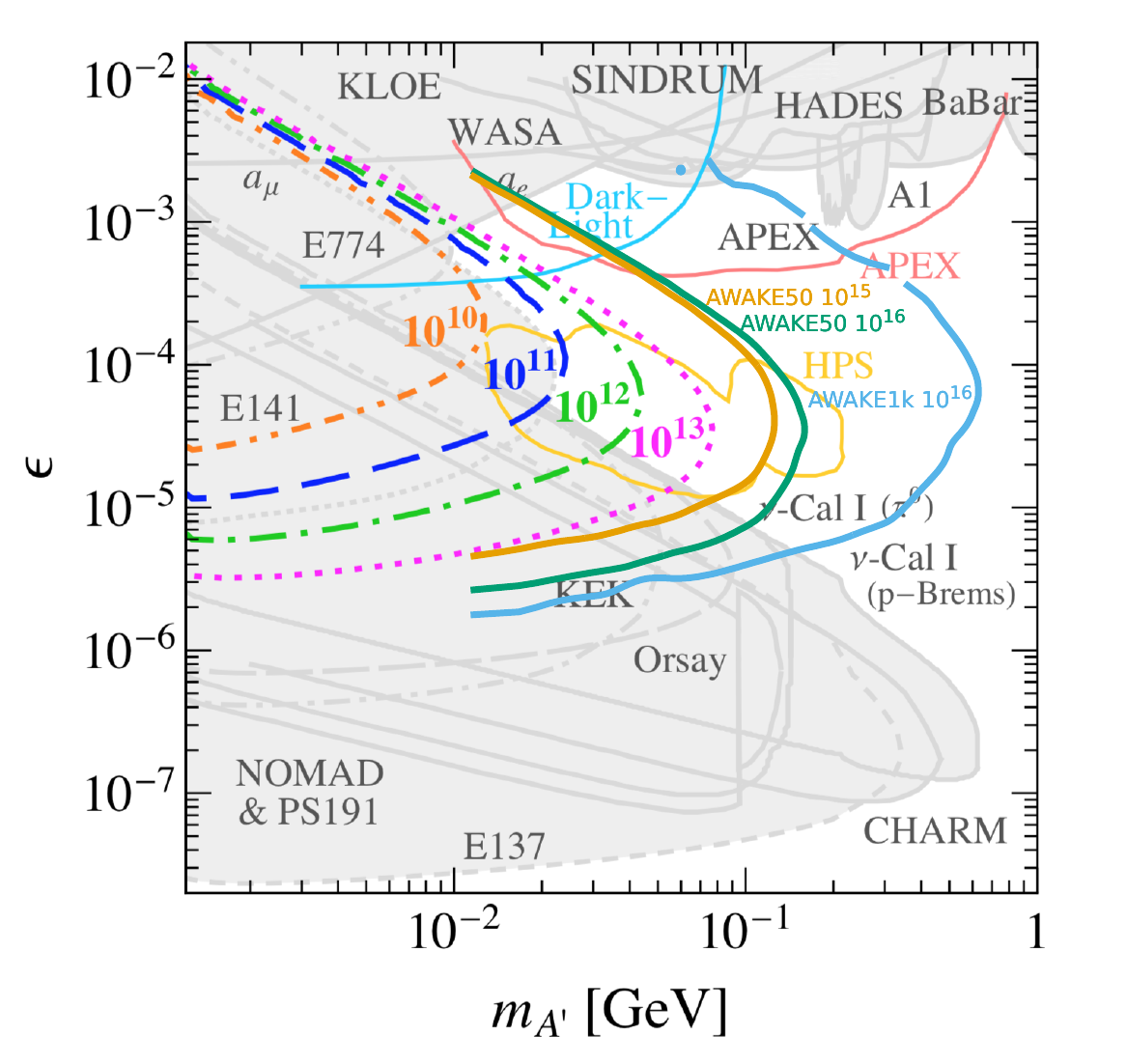}
\caption{Limits on dark photon production decaying to an $e^+e^-$ pair in terms of the mixing strength, $\epsilon$, and dark photon mass, $m_{A^\prime}$, from previous measurements (light grey shading).  The expected sensitivity for the NA64 experiment is shown for a range of electrons on target, $10^{10} - 10^{13}$.  Expectations from other potential experiments are shown as coloured lines.  Expected limits are also shown for $10^{15}$ (orange line) or $10^{16}$ (green line) electrons of 50\,GeV (``AWAKE50'') on target and $10^{16}$ (blue line) electrons of 1\,TeV (``AWAKE1k'') on target provided to an experiment using the future AWAKE accelerator scheme.  From Ref.~\cite{Alemany:2019vsk}.}
\label{fig:dark_photons}
\end{figure}

Expected sensitivities to dark photons in an experiment using an electron beam from the AWAKE scheme are shown in Fig.~\ref{fig:dark_photons} in comparison with previous, current and proposed experiments.  Limits from previous experiments are shown as grey-shaded areas in the $\epsilon-m_{A^\prime}$ plane.  Expected sensitivities from current and future experiments are shown as coloured lines.  Using electrons of energy 50\,GeV will allow dark photon searches to be extended towards higher masses in the range of couplings, $10^{-5} < \epsilon < 10^{-3}$.  As there is the possibility of producing much higher energy electron beams, at the TeV scale, with the AWAKE scheme using the LHC protons as the wakefield driver, the sensitivity is 
shown for 1\,TeV electrons; such an experiment could run as part of a future collider facility, such as the very high energy electron--proton (VHEeP) collider~\cite{epj:c76:463}, after 
collisions, and as the beam is dumped.  The sensitivity is extended to much higher mass values as well as lower $\epsilon$.  
The mass values reached approach 1\,GeV, far beyond the capability of any other experiment, current or planned.

Depending on the future running of the SPS accelerator which feeds 
the AWAKE experiment, a larger area of parameter space could be investigated if more electrons on target were to be possible.  Also, recent investigations indicate that higher energy electrons, up to about 200\,GeV, are possible~\cite{Lotov:2021nob} using the SPS protons as the drive beam which would also extend the sensitivity beyond that shown in Fig.~\ref{fig:dark_photons} for 50\,GeV electrons.  Additionally, other decay channels, such as $A^\prime \to \mu^+ \mu^-$ or $A^\prime \to \pi^+ \pi^-$ could also be considered and the experiment  optimised to be sensitive to these additional channels.

\subsection{Investigation of strong-field QED in electron--laser collisions}
\label{sec:sfqed}

Progress in high-power laser technology has revived the study of strong-field quantum electrodynamics (QED) since the pioneering experiment, E144~\cite{pr:d60:092004}, that investigated this area of physics in electron--laser collisions in the 1990s.  As electrons pass through the intense laser pulse and so experience strong fields (the higher the intensity, the stronger the fields), QED becomes non-linear and experiments mimic the conditions that occur on the surface of neutron stars, at a black hole's event horizon or in atomic physics.

The E144 experiment at SLAC investigated electron--laser collisions with bunches of electrons, each of energy $\sim$\,50\,GeV.  Experiments (E320 at SLAC and LUXE at DESY with the European XFEL) are underway or planned with high-quality electron bunches, with energies in the 10--20\,GeV range~\cite{luxe,e320}. Given the expectation of bunches of electrons, each of energies in the 10s of GeV range, from the AWAKE scheme, experiments investigating strong-field QED are an obvious initial application with the possibility to also have higher electron energies.  As the rate of electron--laser collisions is limited by the roughly 1\,Hz repetition frequency of high-power lasers, high-rate electron bunches are not required.

\subsection{High energy electron--proton/ion colliders}
\label{sec:ep}

The HERA accelerator complex in DESY, Hamburg, has so far provided the only electron--proton collider.  With electrons and protons at  maximal energies of 27.5\,GeV and 920\,GeV, respectively, a centre-of-mass energy of about 318\,GeV was achieved.  An electron--ion collider (EIC)~\cite{eic} is to start operation in BNL in about a decade with lower centre-of-mass energy than HERA, up to about 140\,GeV.  However, it will have significantly higher luminosity, highly polarised beams and variation in the centre of mass energy and ion species, providing a rich physics programme through its great flexibility.  A higher energy electron--proton collider, LHeC (large hadron--electron collider)~\cite{lhec1,lhec2}, has been proposed using electrons of 50\,GeV in collision with LHC protons so yielding centre-of-mass energies just above the TeV scale. The possibility of an LHeC-like collider based on the AWAKE scheme to produce 50\,GeV electrons is outlined here.  A significantly more compact design should be possible, although with a much reduced luminosity performance.  Even more compelling is the possibility of using the AWAKE scheme to provide TeV electrons and so have electron--proton collisions at centre-of-mass energies of 9\,TeV~\cite{epj:c76:463} and this is briefly summarised.  Electrons at 3\,TeV could also be used in fixed-target mode and provide a centre-of-mass energy of $\sim$\,80\,GeV, thereby achieving similar values to the EIC. 

A high energy $ep/eA$ collider could be the first collider application of the AWAKE scheme.  In comparison to a high energy $e^+e^-$ collider, an $ep/eA$ collider poses fewer challenges to the AWAKE scheme as only one beam is required, such low emittances are not needed (as the proton emittance dominates) and potentially positrons are not needed, although they are desirable.  Given the possibility of providing $O(50\,{\rm GeV})$ electrons using the SPS protons to drive wakefields, this was formulated in the PEPIC (plasma electron--proton/ion collider) project.  The electrons would collide with protons from the LHC and using their expected parameters during the high-luminosity phase, this would lead to an instantaneous luminosity of $1.5 \times 10^{27}\,{\rm cm}^{-2}{\rm s}^{-1}$, where parameters expected from a future AWAKE facility have been assumed for the electron bunches~\cite{awake++}.  So although PEPIC has the same energy reach (and possibly even beyond) as the LHeC, it would have a luminosity many orders of magnitude lower.  Such a low luminosity will not allow investigation of the Higgs sector, detailed measurements of electroweak physics or other phenomena that occur at high $Q^2$, where $Q^2$ in the virtuality of the photon emitted by the electron in the $ep/eA$ collision.  However, processes that occur at low $x$, where $x$ is the fraction of the proton's momentum carried by the struck parton, have very high cross sections and so even with a low luminosity, large event samples will be produced.  The focus of the physics programme would then be on understanding the structure of matter and quantum chromodynamics (QCD), the theory of the strong force, in a new kinematic regime. 

The PEPIC collider would be an option for CERN should the LHeC project not go ahead, where the initial study shows that it is possible, could be housed within the current CERN site, and has a novel particle physics programme.  Schemes to increase the luminosity should be considered where the current design is mainly limited by the filling time of the SPS.  Mechanisms to increase the SPS bunch repetition frequency would lead to a corresponding increase in luminosity and so should be studied, along with other parameters relevant for the luminosity such as the electron bunch population and proton bunch size.  The possibility to accelerate electrons up to 200\,GeV using the SPS protons as the drive beam would lead to a doubling of the centre-of-mass energy (2.4\,TeV) and so an increased kinematic range.  This provides a larger phase space to search for new physics and larger lever arm to investigate the energy dependence of high-energy cross sections.

Using the LHC protons to drive wakefields could lead to electrons at the TeV scale and so an $ep$ collider with centre-of-mass energies $O(10\,{\rm TeV})$.  This has been formulated as the VHEeP collider~\cite{epj:c76:463} in which electrons at 3\,TeV are collided with the LHC protons at 7\,TeV giving a centre-of-mass energy of 9\,TeV.  This represents a factor of 30 increase compared to HERA and hence an extension in $x$ and $Q^2$ of a factor of 1000.  As with PEPIC, the luminosity of VHEeP is relatively low, estimated~\cite{vheep-dis2015} to be $10^{28} - 10^{29}\,{\rm cm}^{-2}{\rm s}^{-1}$ or around 1\,pb$^{-1}$ per year.  This is mainly limited by the LHC protons needed to drive wakefields which will need to be dumped after this process and so protons will need to be refilled for the further acceleration of electrons by the AWAKE scheme.  Schemes need to be considered such as squeezing the proton or electron bunches, having multiple interaction points, etc. that will increase the luminosity.

Even though the luminosity of the VHEeP collider is modest, the very high energy provides a compelling particle physics cases.  At low values of $Q^2$, 10s of millions of events are expected and so high precision measurements and searches for new physics will be possible.  This will allow investigation of hadronic cross sections and the structure of matter at very high energies.  The collisions are also equivalent to a photon of energy of 20\,PeV on a fixed target and so has synergy with cosmic-ray physics.  Searches for physics beyond the Standard Model, such as quark substructure or leptoquarks, will also be possible.  Even with modest luminosities, the very high energy ensures that the sensitivity to leptoquarks exceeds that possible in proton--proton collisions at the LHC.  The particle physics case is discussed in more detail elsewhere~\cite{epj:c76:463,Wing:rsta,Caldwell:2018atq}.  As well as $ep$ collisions which have been the focus so far, $eA$ collisions should be considered.

\subsection{Use of BNL proton beams for a compact electron injector for a future electron--ion collider}
\label{sec:eic}

The AWAKE acceleration scheme has been considered as a possible technology for the injector for the high energy electron beam 
for the future EIC.  The EIC is expected to collide electrons of up to 20\,GeV with protons of up to 275\,GeV -- 
a possible site at BNL in the US already has a circular, 4\,km long proton accelerator, but will require a 
new electron accelerator of similar size.  It has been proposed~\cite{Chappell:2019ovd} to use the AWAKE technology at BNL 
by using their high-intensity proton bunches to generate large wakefields and hence accelerate electrons to high energies over short 
distances.  Along with CERN, where AWAKE is currently based, BNL is one of the few places in the world with 
high energy proton bunches that can be used for the AWAKE acceleration scheme.

\begin{figure}[thp!]
\centering
\includegraphics[width=0.6\textwidth]{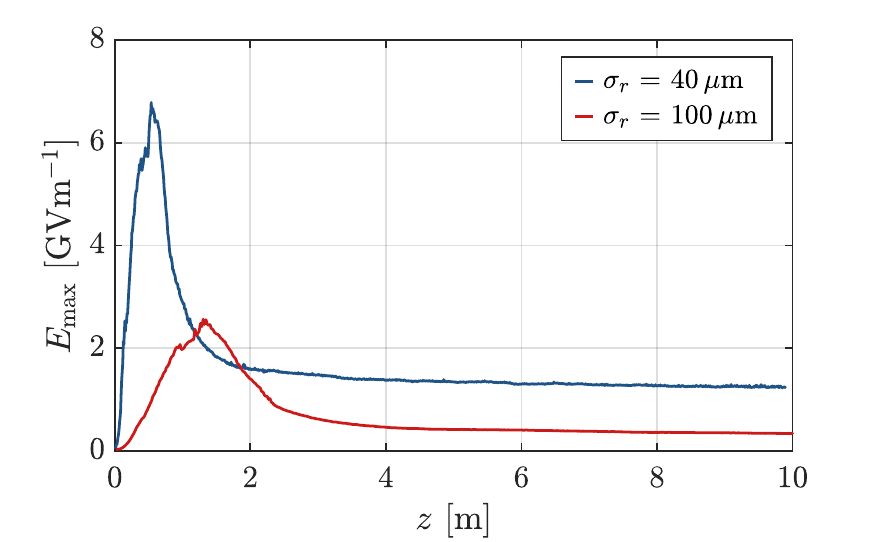}
\caption{
    Evolution of the peak longitudinal fields driven by the BNL proton drive beams over 10\,m using bunch parameters which differ only in their transverse size, $\sigma_r = 40\,\mu$m or $100\,\mu$m. 
    From Ref.~\cite{Chappell:2019ovd}.}
\label{fig:eic}
\end{figure}

It has been assumed that the proton bunch parameters expected for the future EIC, namely a proton energy of 275\,GeV, 
$2 \times 10^{11}$\,protons/bunch, a bunch length of 5\,cm and a bunch radius of $40$ or $100\,\mu$m.  These are similar values to currently 
used at AWAKE using CERN's SPS accelerator except that the smaller bunch size allows for a higher plasma density to be used 
which should yield larger accelerating fields.  Using these parameters, the process has been simulated and the accelerating electric 
field determined for the two values of the bunch radius.  The results are shown in Fig.~\ref{fig:eic}, where the larger beam radius 
is given by the red line and smaller beam radius by the blue line.  In the more promising scenario, with a bunch radius of $40\,\mu$m, 
the peak field is almost 7\,GV/m, which although falls rapidly, levels off above 1\,GV/m, an accelerating gradient well above current 
conventional accelerator technology.

This 
confirms that the high, $> 1$\,GV/m gradients, can be harnessed and that the required maximum energy for the EIC of 20\,GeV could 
be achieved in under 20\,m.  There are indications that the accelerating gradient can be frozen~\cite{PoP22-103110} soon after the peak field, which will be studied at AWAKE (see Sections~\ref{sec:density-step} and~\ref{sec:density-step2}), 
and values more like $5-6$\,GV/m would be observed in Fig.~\ref{fig:eic} in which case electrons could be accelerated to the 
required 20\,GeV in under 4\,m.  Alternatively, 
significantly higher electron energies than currently planned, of 50\,GeV and beyond, could be achieved and so extend the kinematic reach in the investigation of parton dynamics in the proton at the EIC.

\section{Summary}
\label{sec:summary}

This article summarised the plans for AWAKE Run\,2, which will take the scheme of proton-driven plasma wakefield acceleration from proof of concept to a technology which can provide beams for particle physics experiments.  

The proof of concept, AWAKE Run\,1, showed that a long proton bunch can be modulated into a series of microbunches, which are regularly and reproducibly spaced.  These microbunches constructively interfere to generate strong electric fields in their wake.  The wakefields were sampled by an externally-injected bunch of electrons which were accelerated from about 20\,MeV up to about 2\,GeV within 10\,m of plasma, representing an average field of $> 200$\,MV/m, with peak fields of up to $\sim 1$\,GV/m expected.

In AWAKE Run\,2, the proof of concept will be significantly extended to address the requirements needed to develop beams for use in particle physics experiments.  The primary aims of AWAKE Run\,2 are to sustain the expected peak fields of $0.5 - 1$\,GV/m over long distances, thereby increasing the accelerated electron energy; to demonstrate that the emittance of the electron bunch is preserved during acceleration in plasma; and to develop plasma sources that are scalable to 100s of metres and beyond.  This will be achieved in a staged approach during the 2020s which will require significant extension to the current AWAKE facility, in particular the development of short witness electron bunches for injection, new plasma sources and a suite of diagnostics to measure the physics of the acceleration process.

After completion of AWAKE Run\,2, at the end of the 2020s, the scheme should have been sufficiently demonstrated such that it can be used to provide beams for particle physics experiments.  Given the challenge in producing high energy electron bunches by conventional means, electrons in the 20--200 GeV range, as driven by protons from the SPS, or even at the TeV scale if using LHC protons as the wakefield driver, can be used in a variety of particle physics experiments.  Such electron bunches can be used in experiments to search for dark photons, to measure QED in strong fields or as the injector or main accelerator for the electron arm of an electron--proton or electron--ion collider.  These first applications place less stringent requirements on the parameters of the electron bunch than for a high energy, high luminosity linear electron--positron collider, although they will provide a useful stepping stone, along with continued R\&D, to such ultimate applications whilst also providing beams for novel particle physics experiments.

\authorcontributions{
E.G, K.L., P.M and M.W. conceived, wrote and edited this article.  Members of the AWAKE collaboration contributed to the review of the manuscript and to the work here summarised.}

\funding{
This work was supported in parts by a Leverhulme Trust Research Project Grant RPG-2017-143 and by STFC (AWAKE-UK, Cockcroft Institute core, John Adams Institute core, and UCL consolidated grants), United Kingdom;
the Russian Science Foundation, project 20-12-00062, for Novosibirsk's contribution;
the National Research Foundation of Korea (Nos.\ NRF-2016R1A5A1013277 and NRF-2020R1A2C1010835);
the Wolfgang Gentner Programme of the German Federal Ministry of Education and Research (grant no.\ 05E15CHA);
M. Wing acknowledges the support of DESY, Hamburg.
Support of the National Office for Research, Development and Innovation
(NKFIH) under contract numbers 2019-2.1.6-NEMZ\_KI-2019-00004 and
MEC\_R-140947, and the support of the Wigner Datacenter Cloud facility
through the Awakelaser project is acknowledged.
The work of V. Hafych has been supported by the European Union's Framework Programme for Research and Innovation Horizon 2020 (2014--2020) under the Marie Sklodowska-Curie Grant Agreement No.\ 765710.
TRIUMF contribution is supported by NSERC of Canada.
}



\dataavailability{
No new data is provided in this article which presents a review of previous work.}

\acknowledgments{
The AWAKE collaboration acknowledge the SPS team for their excellent proton delivery.}

\conflictsofinterest{
The authors declare no conflict of interest.}

\abbreviations{The following abbreviations are used in this manuscript:\\

\noindent 
\begin{tabular}{@{}ll}
AWAKE & Advanced wakefield experiment\\
BNL & Brookhaven National Laboratory\\
CERN & European Organisation for Nuclear Research \\ 
& (Conseil Europ\'{e}en pour la Recherche Nucl\'{e}aire)\\
CNGS & CERN neutrinos to Gran Sasso\\
DESY & Deutsches Elektronen-Synchrotron\\
EIC & Electron--ion collider\\
GEANT & Geometry and tracking\\
HERA & Hadron--electron ring accelerator\\
LHC & Large hadron collider\\
LHeC & Large hadron--electron collider\\
LUXE & Laser und XFEL experiment\\
PEPIC & Plasma electron--proton/ion collider\\
QCD & Quantum chromodynamics\\
QED & Quantum electrodynamics\\
RIF & Relativistic ionisation front\\
RMS & Root mean square\\
SLAC & Stanford Linear Accelerator\\
SM & Self-modulation\\
SMI & Self-modulation instability\\
SPS & Super proton synchrotron\\
VHEeP & Very high energy electron--proton\\
XFEL & X-ray free electron laser
\end{tabular}}



\end{paracol}
\reftitle{References}

\end{document}